\documentclass[amsmath,amssymb,aps,prd,reprint,twocolumn,showkeys,showpacs]{revtex4-1}

\usepackage[utf8]{inputenc}
\usepackage{amsmath}
\usepackage{graphicx}
\usepackage{dcolumn}
\usepackage{bm}
\usepackage{epstopdf}
\printfigures
\usepackage{epsfig}
\usepackage{mathtools}
\usepackage{subfigure}
\usepackage{subfig}

\usepackage{booktabs}

\begin{document}

\title{Dynamical description of a quintom cosmological model nonminimally coupled with gravity}

\author{Mihai Marciu}
\email{mihai.marciu@drd.unibuc.ro}
\affiliation{Faculty of Physics, University of Bucharest, 405 Atomi\c{s}tilor, POB MG-11, RO-077125, Bucharest-M\u{a}gurele, Romania}

\date{\today}

\begin{abstract}
In this work we have studied a cosmological model based on a quintom dark energy model non--minimally coupled with gravity, endowed with a specific potential energy of the exponential squared type. For this specific type of potential energy and non--minimal coupling, the dynamical properties are analyzed and the corresponding cosmological effects are discussed. Considering the linear stability method, we have investigated the dynamical properties of the phase space structure, determining the physically acceptable solutions. The analysis showed that in this model we can have various cosmological epochs, corresponding to radiation, matter domination, and de Sitter eras. Each solution is investigated from a physical and cosmological point of view, obtaining possible constraints of the model's parameters. In principle the present cosmological setup represent a possible viable scalar tensor theory which can explain various transitional effects related to the behavior of the dark energy equation of state and the evolution of the Universe at large scales.

\end{abstract}

\maketitle

\section{Introduction}
\par 
In the present days the cosmological context reached a golden age epoch by overturning the fundamental concepts related to the evolution and the major constituents of the Universe, fracturing our understanding of time and space. The accelerated expansion of the Universe \cite{Huterer:2017buf} represent an enigma to theorists and cosmologists, with deep ramifications in various branches of physics. The basic evidence of the accelerated expansion has been probed through various astrophysical studies \cite{Jones:2017udy,Abbott:2018wzc} which included observation from type Ia supernovae \cite{Macaulay:2018fxi,Hicken_2009}, baryon acoustic oscillations \cite{V_liviita_2015, Ryan:2018aif,Abbott:2017wcz,Chan:2018pjr,Avila:2017nyy} and cosmic microwave background radiation \cite{Pourtsidou:2016ico, Murgia:2016ccp, Wang:2018ahw}. The simplest scenario of dark energy is represented by the cosmological constant \cite{Copeland:2006wr} added to the Einstein field equation, a proposal which lead to a constant equation of state for the dark energy sector. In order to explain the dynamical evolution of the dark energy equation of state various theoretical directions have been proposed \cite{Silvestri:2009hh, Bamba:2012cp,Elizalde:2004mq,Odintsov:2018qyy, Nojiri:2005sx, Nojiri:2003vn,Elizalde:2008yf} in the form of single or multiple scalar fields, minimally or non--minimally coupled with gravity or other possible invariants \cite{Bahamonde:2017ize}.
\par 
The behavior of the dark energy equation of state \cite{Melchiorri:2002ux,Joyce:2016vqv,PhysRevLett.109.171301} represents an important aspect when constructing a viable scalar tensor theory of gravitation \cite{Mukaigawa:1997nh, Odintsov:1990mt} which can explain various physical quantities associated to the known Universe. In this case the strange issue related to the crossing over the phantom divide line (the cosmological constant barrier) \cite{PhysRevLett.109.171301,Zhao:2017cud} by the dark energy equation of state has been explained by adopting a possible extension to the Einstein--Hilbert action which adds two scalar fields \cite{Feng:2004ad,Guo:2004fq}, an addition which includes a canonical scalar field and a phantom field, respectively, a composition which violates the null energy condition \cite{Cai:2009zp}. In scalar tensor theories the quintessence dark energy models \cite{Tsujikawa:2013fta, Zlatev:1998tr} represent a possible configuration for the dark energy sector, a canonical direction which can explain various astrophysical observations. A more exotic configuration which includes the addition of a negative kinetic energy in the specific action has been suggested \cite{PhysRevLett.91.071301, Caldwell:1999ew,PhysRevD.68.023509}, leading to the formation of phantom dark energy models, a particular theoretical direction which is viable from an observational point of view \cite{Barboza:2008rh,Ludwick:2017tox, Zhao:2006bt}. However, such theoretical constructions lead to the violation of the null energy condition \cite{PhysRevD.68.023509,  Easson:2016klq, Sawicki:2012pz} and can exhibit Big Rip ending scenes. Since the nature of the dark energy section is currently unknown, various exotic models have been proposed \cite{Bahamonde:2017ize}, adding new intriguing directions to the cosmic landscape. 
\par 
The development of the scalar tensor theories lead to the formation of quintom cosmological models \cite{Cai:2009zp}, an exotic configuration which might explain some of the dynamical aspects associated to the dark energy equation of state. In the first quintom scenario the two quintom scalar fields were minimally coupled in the corresponding action \cite{Feng:2004ad,Guo:2004fq}, explaining the astrophysical observations related to the specific crossing \cite{Feng:2004ff} of the cosmological constant boundary. In the recent years the quintom paradigm \cite{Cai:2009zp} has been continuously developed in various studies \cite{Chimento:2008ws,Panpanich:2019fxq,Sadeghi:2019btx, Leon:2018lnd, Sadeghi:2017nxe, MohseniSadjadi:2006hb, Lazkoz:2007mx,Dutta:2016exd,  Setare:2008sf, Shi:2008df, Saridakis:2009ej,Saridakis:2009jq, Qiu:2010ux, Amani:2011zz, Leon:2012vt} which includes the additions of various non--minimal couplings in different scalar tensor theories \cite{Setare:2009mc, PhysRevD.93.123006, Bahamonde:2018miw, Behrouz:2017sqh, Marciu:2018oks, Marciu:2019cpb,Deffayet:2010qz}. In spite of the fact that the quintom paradigm implies the violation of the null energy condition, embedding a pathological phantom field in the corresponding action, it remains as an admissible modified gravity construction which can justify various astrophysical observations \cite{Cai:2009zp}. Although a quintom action based on two scalar fields include the addition of a phantom field which lead to specific instabilities when possible quantum features are considered, it is consistent with astrophysical observations, showing the specific effect related to the crossing of the phantom divide line by the dark energy equation of state, a dynamical effect \cite{Cai:2009zp} which cannot be explained in single scalar field models with minimal coupling \cite{Vikman:2004dc}. In an earlier paper \cite{PhysRevD.93.123006}, a quintom dark energy extension has been proposed, where the scalar fields were non--minimally coupled with scalar curvature, the physical features of the model were analyzed by adopting a numerical approach. In scalar tensor theories of gravitation the addition of non--minimal couplings with gravity represent a viable direction supported by different hypothetical models \cite{Chernikov:1968zm,PhysRevD.35.2955,PhysRevD.62.023504, PhysRevLett.42.417, Faraoni:2000gx, Sonego_1993}. The effects of the non--minimal couplings with gravity have been investigated in single scalar field theories \cite{Hrycyna:2015vvs, Hrycyna:2009zf,Kerachian:2019tar,Uzan:1999ch}, by considering the linear stability theory, showing the viability of the corresponding models \cite{Hrycyna:2007gd, Hrycyna:2015eta, Szydlowski:2008zza, Szydlowski:2008in,HRYCYNA2010191, Hrycyna:2010yv, Hrycyna:2008gk, Szydlowski:2013sma}. Furthermore, in scalar tensor theories based on teleparallel gravity the models non--minimally coupled with gravity are constructed by using the corresponding analogous invariant scalars, the torsion \cite{Geng:2011aj} and boundary coupling \cite{Bahamonde:2015hza} parameters. From an observational point of view the non--minimal couplings with curvature have been investigated in different specific models \cite{Hrycyna:2015vvs,Luo:2005ra, Nozari:2007eq, Szydlowski:2008zza}. The extension of the quintom paradigm towards non—minimal curvature couplings represents a particular attempt of correcting two scalar field models, a specific model which might explain the dynamical crossing \cite{Zhao:2017cud,Cai:2009zp} of the cosmological constant boundary in the recent past by the dark energy equation of state, a phenomenon presented by recent astrophysical observations.
\par 
In this paper we shall further analyze the dynamical features of a specific scalar tensor cosmological scenario \cite{PhysRevD.93.123006}, observing the physical consequences of the couplings between the quintom scalar fields and the curvature in the phase space, for a different potential energy, considering the linear stability method. The potential energy type considered in the present paper belongs to the exponential squared class, which have been previously studied \cite{HRYCYNA2010191} in scalar tensor theories of gravitation.
\par 
The paper is organized as follows: in Sec.~II we present the basic equations which express the evolution relations for the quintom model non--minimally coupled to scalar curvature, endowed with a specific potential energy of exponential squared type. Then, in Sec.~III we propose the auxiliary variables and write the autonomous system of equations, determining the critical points and the dynamical features which are associated. In the last section Sec.~IV, we present the summary of the analytical  investigation and the final concluding remarks.

\section{The Field equations and modified Friedmann relations }
\label{sec:doi}
In what follows we shall study a quintom model for the dark energy component non--minimally coupled with scalar curvature, which includes an action corresponding to the matter component $S_m$, assuming the following form of the total action \cite{PhysRevD.93.123006}:

\begin{multline}
\label{actiune}
S_{tot}=S_m+\int d^4x \sqrt{-g}\frac{R}{2}+\frac{1}{2}\int d^4x \sqrt{-g} \Big(-g^{\mu\nu}\partial_{\mu}\phi \partial_{\nu}\phi
\\+g^{\mu\nu}\partial_{\mu}\sigma \partial_{\nu}\sigma
-\xi_1 R \phi^2+\xi_2 R \sigma^2 -2 V_1(\phi)-2 V_2(\sigma) \Big),
\end{multline}
where $\phi(t)$ represents the canonical scalar field (quintessence), and $\sigma(t)$ the non--canonical (negative kinetic) field with a phantom pathological behavior; $R$ denotes the scalar curvature which for the metric descriptor $(-1, +a^2(t),+a^2(t), +a^2(t))$ is equal to $R=6(\frac{\ddot{a}}{a}+(\frac{\dot{a}}{a})^2)$. Here we shall assume the fields $\phi$ and $\sigma$ to be time dependent, and use dots to denote derivatives with respect to the cosmic time. Also, $a(t)$ is the usual cosmic scale factor and $H=\dot{a}/a$ the associated Hubble parameter.
\par 
The modified Friedmann relations for this specific action are the following \cite{PhysRevD.93.123006}:
\begin{equation}
3 H^2=\rho_{\phi}+\rho_{\sigma}+\rho_m,
\end{equation}
\begin{equation}
\dot{H}=-\frac{1}{2}(\rho_{\phi}+\rho_{\sigma}+\rho_m+p_{\phi}+p_{\sigma}+p_m),
\end{equation}
with the corresponding energy densities and pressures \cite{HRYCYNA2010191}:
\begin{equation}
\rho_\phi=\frac{1}{2}\dot{\phi}^2+V_1(\phi)+3 \xi_1 H^2 \phi^2+6 \xi_1 H \phi \dot{\phi},
\end{equation} 
\begin{equation}
\rho_\sigma=-\frac{1}{2}\dot{\sigma}^2+V_2(\sigma)-3 \xi_2 H^2 \sigma^2-6 \xi_2 H \sigma \dot{\sigma},
\end{equation}
\begin{equation}
p_{\phi}=\frac{1}{2}\dot{\phi}^2-V_1(\phi)-\xi_1 (\phi^2(3 H^2+2 \dot{H})+2 \phi \ddot{\phi}+2 \dot{\phi}^2+4 H \phi \dot{\phi}),
\end{equation} 
\begin{equation}
p_{\sigma}=-\frac{1}{2}\dot{\sigma}^2-V_2(\sigma)+\xi_2 (\sigma^2(3 H^2+2 \dot{H})+2 \sigma \ddot{\sigma}+2 \dot{\sigma}^2+4 H \sigma \dot{\sigma}).
\end{equation} 
\par 
Furthermore, we can define the pressure, 
\begin{equation}
p_{\phi\sigma}=p_{\phi}+p_{\sigma},
\end{equation} 
the energy density for the dark energy component,
\begin{equation}
\rho_{\phi\sigma}=\rho_{\phi}+\rho_{\sigma},
\end{equation} 
the dark energy equation of state
\begin{equation}
w_{\phi\sigma}=\frac{p_{\phi}+p_{\sigma}}{\rho_{\phi}+\rho_{\sigma}},
\end{equation}
and the effective (total) equation of state:
\begin{equation}
w_{\bf{eff}}=\frac{p_m+p_{\phi \sigma}}{\rho_m+\rho_{\phi \sigma}}=\frac{p_m+p_{\phi}+p_{\sigma}}{\rho_m+\rho_{\phi}+\rho_{\sigma}}.
\end{equation}
\par 
In this case we define the matter and dark energy energy density parameters
\begin{equation}
\Omega_m=\frac{\rho_m}{3 H^2}, 
\end{equation}
\begin{equation}
\Omega_{\phi\sigma}=\frac{\rho_\phi+\rho_{\sigma}}{3 H^2}, 
\end{equation}
which will obey the constraint equation:
\begin{equation}
\Omega_m+\Omega_{\phi\sigma}=1.
\end{equation}
\par 
Next, for the present cosmological model we have obtained the following Klein--Gordon relations from the principle of least action \cite{PhysRevD.93.123006,HRYCYNA2010191}:
\begin{equation}
\ddot{\phi}+3 H \dot{\phi}+\xi_1 R \phi+\frac{d V_1 (\phi)}{d \phi}=0,
\end{equation}
\begin{equation}
\ddot{\sigma}+3 H \dot{\sigma}+\xi_2 R \sigma-\frac{d V_2 (\sigma)}{d \sigma}=0.
\end{equation}
\par 
Considering the above relations, it can be shown that the dark energy field obeys a standard continuity equation:
\begin{equation}
\dot{\rho_{\phi\sigma}}+3 H (\rho_{\phi \sigma}+ p_{\phi\sigma})=0.
\end{equation} 
\par 

\onecolumngrid

\begin{table}[h!]
\centering
 \begin{tabular}{||c | c |c| c| c| c| c| c | c | c||} 
 \hline
 Point & $x_1$ & $x_2$ & $y_1$ & $y_2$ & $z_1$ & $z_2$ & $\Omega_m $ & $\Omega_{\rho\sigma}$ &  $w_{\bf{eff}}$ \\  
 \hline\hline
 $P_{1-} $ & 0 & 0 & 0 & 0 & $-\frac{\sqrt{1+6 \xi_2 z_2^2}}{\sqrt{6 \xi_1}}$ & $z_2$ & 0 & 1 & $\frac{1}{3}$\\
  \hline
   $P_{1+}$ & 0 & 0 & 0 & 0 & $+\frac{\sqrt{1+6 \xi_2 z_2^2}}{\sqrt{6 \xi_1}}$ & $z_2$ & 0 & 1 & $\frac{1}{3}$\\
    \hline
  $P_{2-}$ & 0 & 0 & $\frac{2 \sqrt{6 \xi_1}}{\sqrt{\alpha_1}}$ & $2\sqrt{-6\frac{\xi_2}{\alpha_2}}$ &   $-\frac{\sqrt{-24 \alpha_2 \xi_1 + \alpha_1 (\alpha_2+ 24 \xi_2) + 6 \alpha_1 \alpha_2 \xi_2 z_2^2}}{\sqrt{6 \alpha_1 \alpha_2 \xi_1}}$ & $z_2$ & 0 & 1 & -1\\
  \hline
  $P_{2+}$ & 0 & 0 & $\frac{2 \sqrt{6 \xi_1}}{\sqrt{\alpha_1}}$ & $2\sqrt{-6\frac{\xi_2}{\alpha_2}}$ &   $+\frac{\sqrt{-24 \alpha_2 \xi_1 + \alpha_1 (\alpha_2+ 24 \xi_2) + 6 \alpha_1 \alpha_2 \xi_2 z_2^2}}{\sqrt{6 \alpha_1 \alpha_2 \xi_1}}$ & $z_2$ & 0 & 1 & -1\\
    \hline
  $P_{3}$ & 0 & 0 & $y_1$ & $\sqrt{1-y_1^2}$ & 0 & 0 & 0 & 1 & -1\\
  \hline
  $P_{4}$ & 0 & 0 & 0 & 0 & 0 & 0 & 1 & 0 & $w_m$\\
   \hline
  $P_{5-}$ & 0 & 0 & 0 & 0 & $-\frac{1}{\sqrt{6 \xi_1}}$ & 0 & 0 & 1 & $\frac{1}{3}$\\
  \hline
  $P_{5+}$ & 0 & 0 & 0 & 0 & $+\frac{1}{\sqrt{6 \xi_1}}$ & 0 & 0 & 1 & $\frac{1}{3}$\\
  \hline
  $P_{6-}$ & 0 & 0 & 0 & $2\sqrt{6} \sqrt{-\frac{\xi_2}{\alpha_2}}$ & 0 & $-\frac{\sqrt{-\alpha_2-24 \xi_2}}{\sqrt{6 \alpha_2 \xi_2}}$ & 0 & 1 & -1\\
  \hline
  $P_{6+}$ & 0 & 0 & 0 & $2\sqrt{6} \sqrt{-\frac{\xi_2}{\alpha_2}}$ & 0 & $+\frac{\sqrt{-\alpha_2-24 \xi_2}}{\sqrt{6 \alpha_2 \xi_2}}$ & 0 & 1 & -1\\
  \hline
  $P_{7-}$ & 0 & 0 & $\frac{2 \sqrt{ 6 \xi_1}}{\sqrt{\alpha_1}}$ & 0 &  $-\frac{\sqrt{\alpha_1-24 \xi_1}}{\sqrt{6 \alpha_1 \xi_1}}$ & 0 & 0 & 1 & -1\\
  \hline
  $P_{7+}$ & 0 & 0 & $\frac{2 \sqrt{ 6 \xi_1}}{\sqrt{\alpha_1}}$ & 0 &  $+\frac{\sqrt{\alpha_1-24 \xi_1}}{\sqrt{6 \alpha_1 \xi_1}}$ & 0 & 0 & 1 & -1\\
   \hline
  $P_{8}$ & 0 & 0 & $\frac{2 \sqrt{ 6 \xi_1}}{\sqrt{\alpha_1}}$ & $\frac{\sqrt{\alpha_1-24 \xi_1 - 6 \alpha_1 \xi_1 z_1^2}}{\sqrt{\alpha_1}}$ & $z_1$ & 0 & 0 & 1 & -1\\
  \hline
  $P_{9}$ & 0 & 0 & $\frac{\sqrt{\alpha_2+24 \xi_2 + 6 \alpha_2 \xi_2 z_2^2}}{\sqrt{\alpha_2}}$ & $\frac{2 \sqrt{6\xi_2}}{\sqrt{- \alpha_2}}$ & 0 & $z_2$ & 0 & 1 & -1\\
  \hline
 \end{tabular}
 \caption{The location of the critical points and the corresponding physical features.}
\label{table:1}
\end{table}

\twocolumngrid

\begin{figure}[htp]
    \includegraphics[width=0.7\columnwidth]{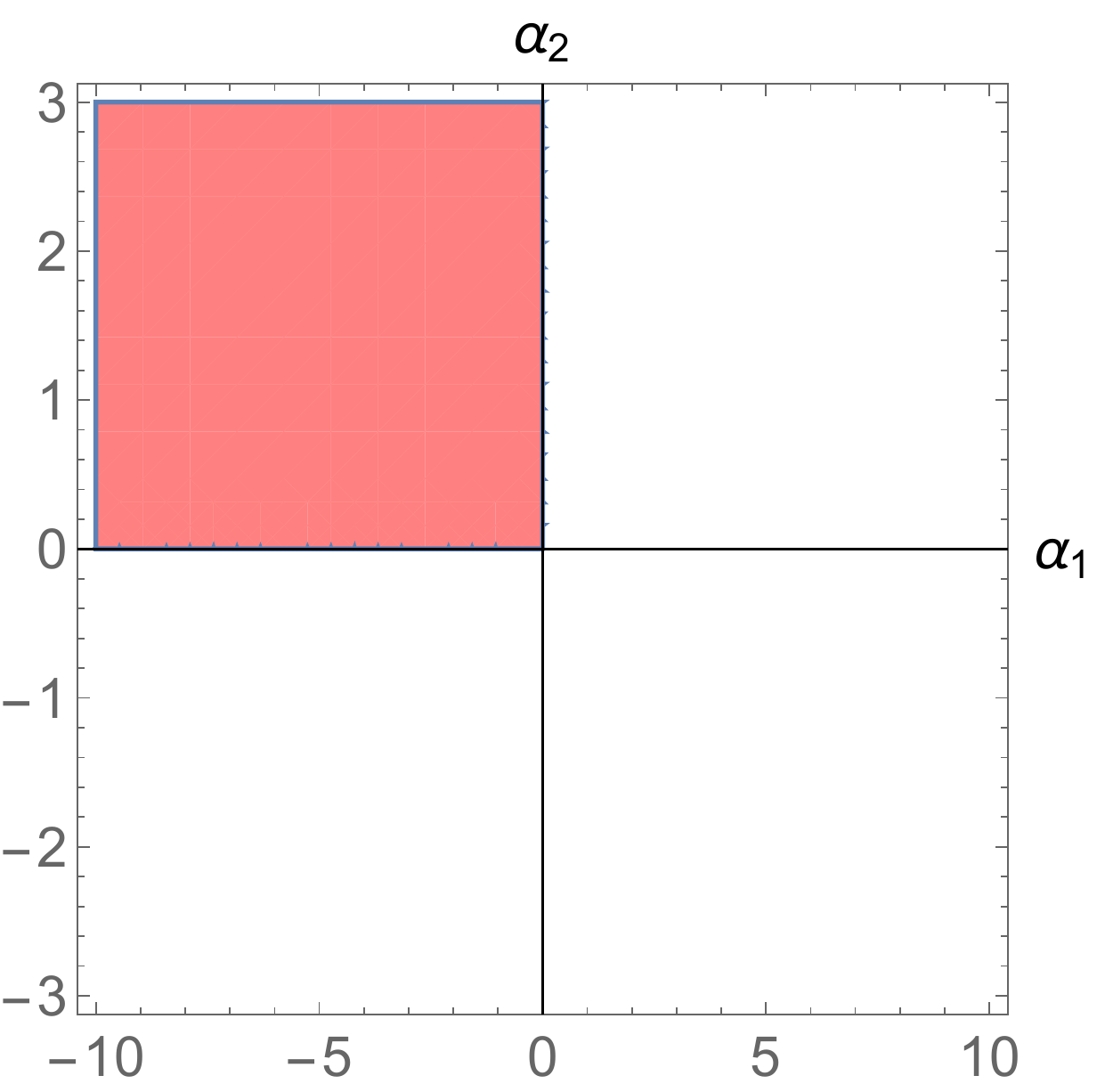}
    \caption{The non-exclusive existence regions for the $P_{2+}$ critical line ($\xi_1=-4, \xi_2=-1, z_2=10, \alpha_1 \in [-10,+10]$).}
    \label{fig1}
\end{figure}  

\begin{figure}[htp]
    \includegraphics[width=0.7\columnwidth]{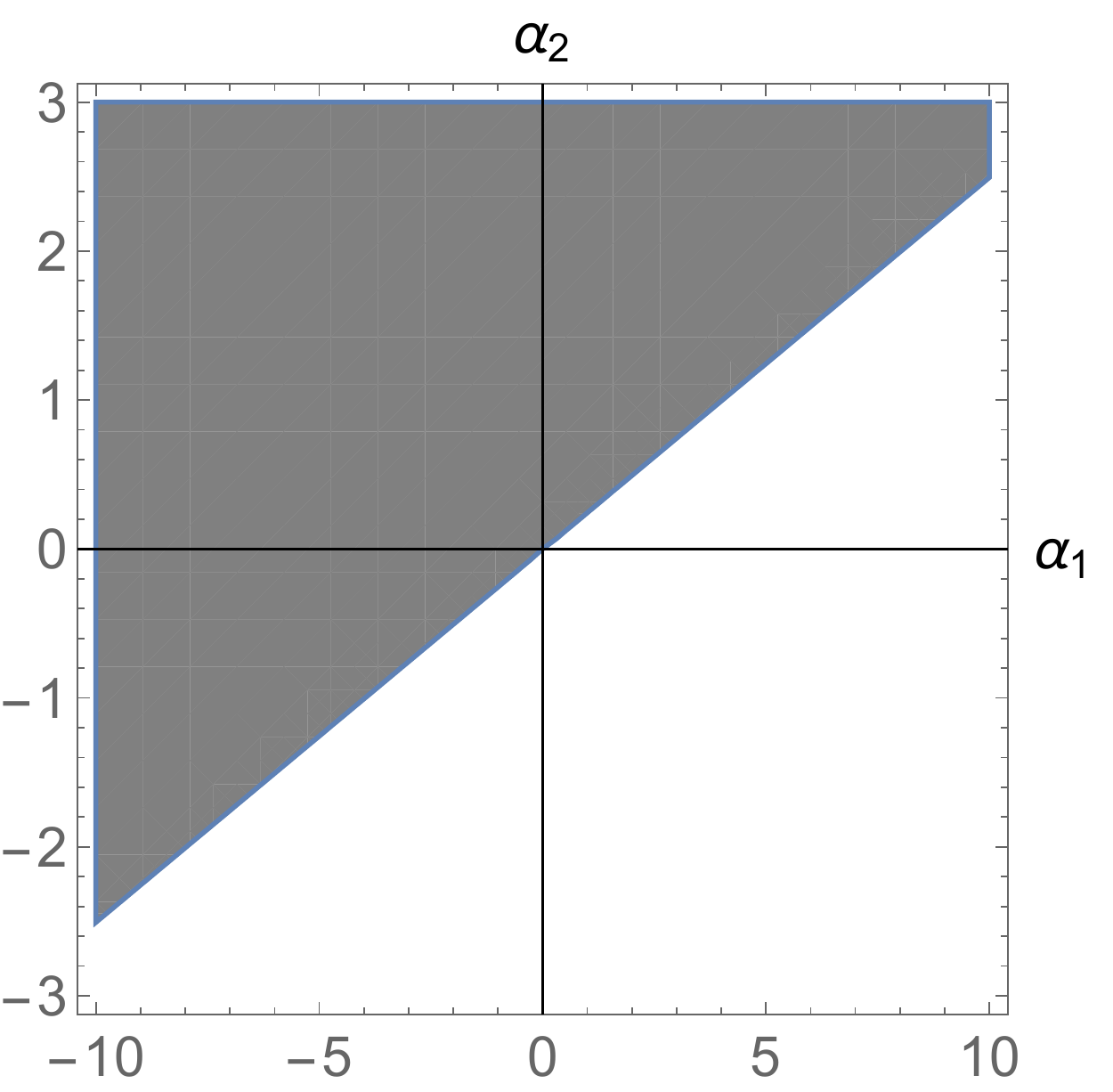}
    \caption{The non-exclusive saddle regions for the $P_{2+}$ critical line ($\xi_1=-4, \xi_2=-1, z_2=10, w_m=0, \alpha_1 \in [-10,+10]$).}
    \label{fig2}
\end{figure}  

\begin{figure}[htp]
    \includegraphics[width=0.7\columnwidth]{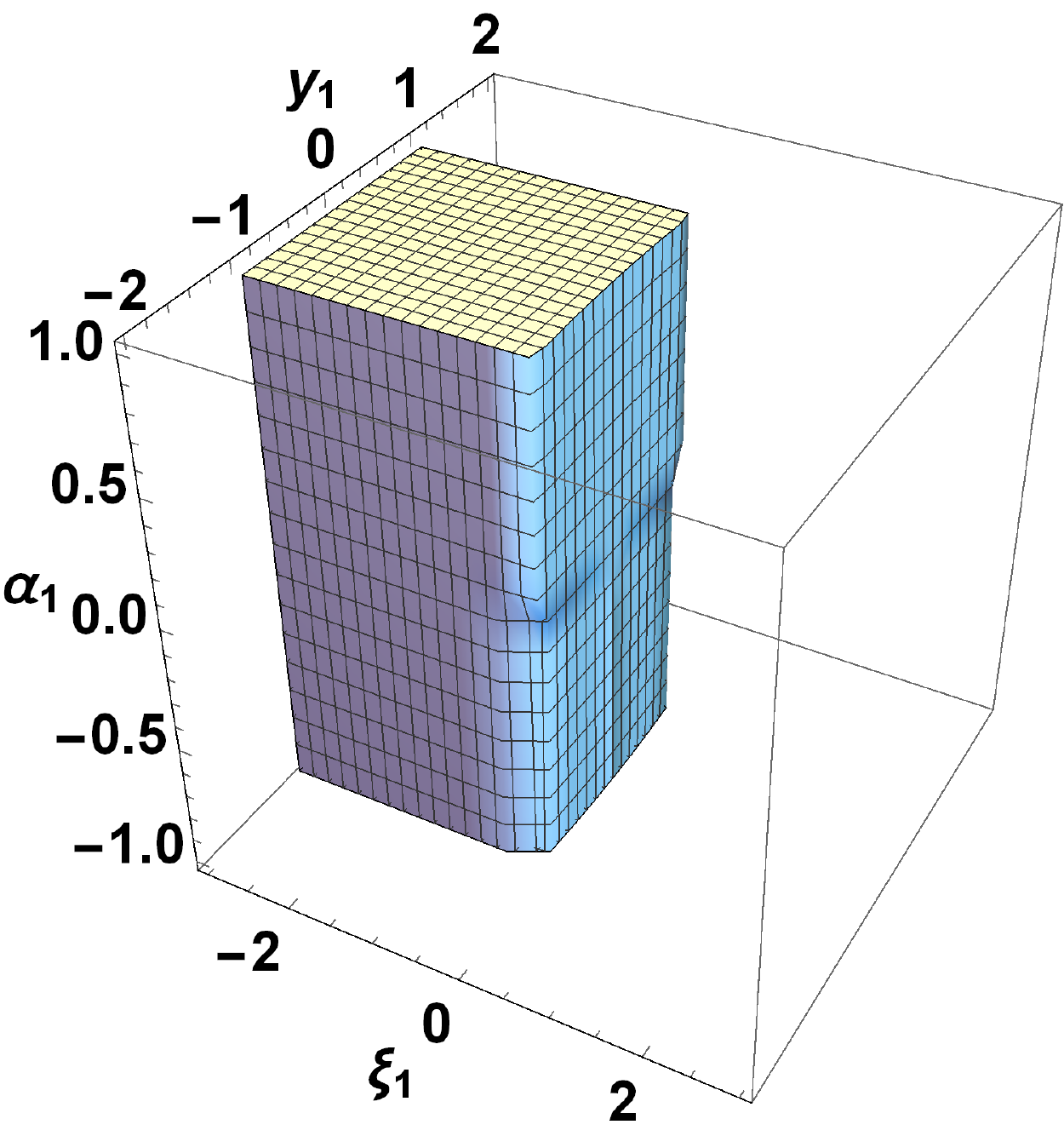}
    \caption{The non-exclusive saddle regions where the $P_{3}$ critical line represents a saddle cosmological epoch ($E_{P_{3}}[3]<0, E_{P_{3}}[4]>0$).}
    \label{fig3}
\end{figure}  

\begin{figure}[htp]
    \includegraphics[width=0.7\columnwidth]{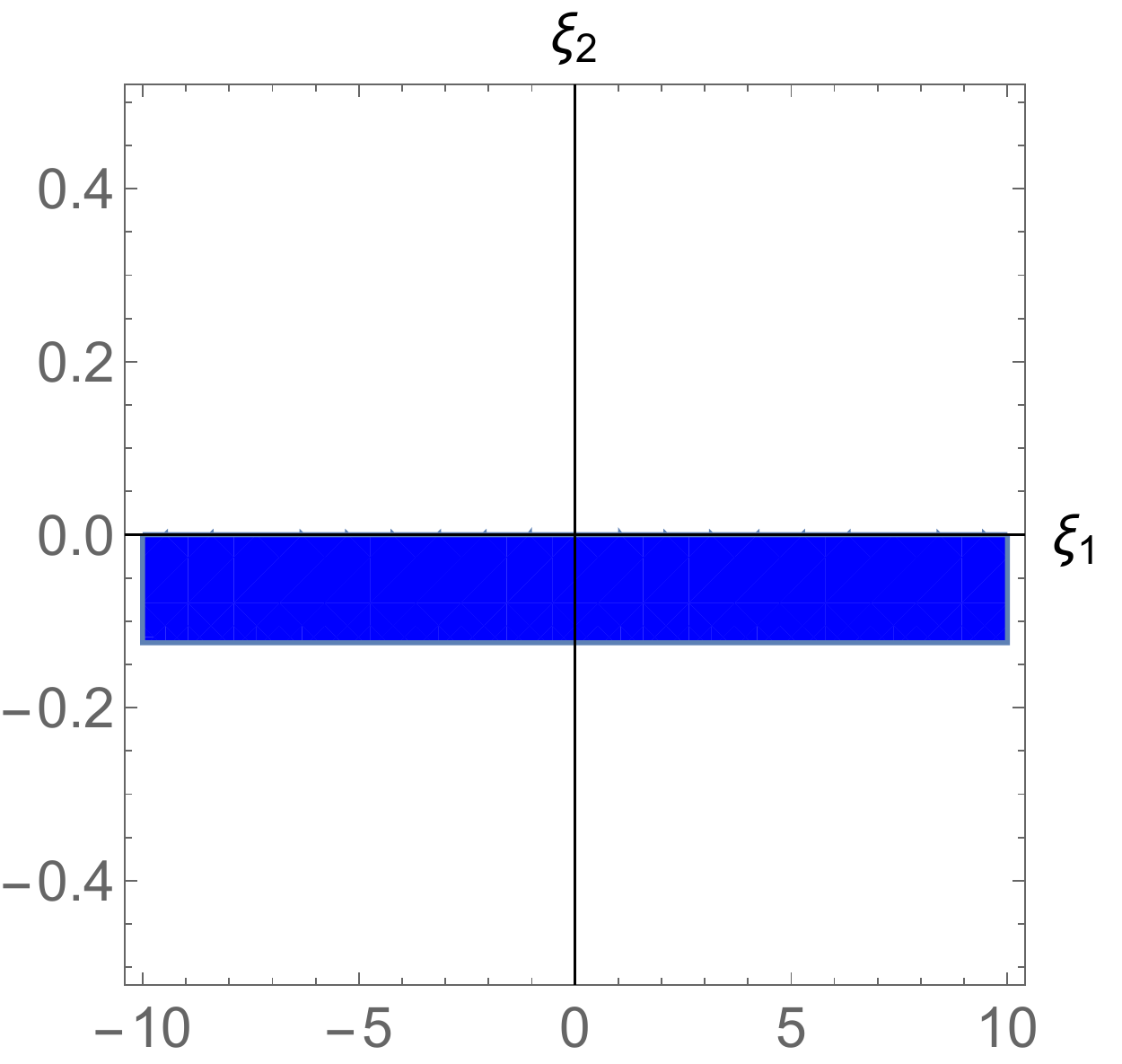}
    \caption{The non-exclusive saddle regions where the $P_{6+}$ critical point represents a saddle cosmological epoch ($\alpha_1=1, \alpha_2=3, w_m=0, \xi_1 \in [-10,+10]$).}
    \label{fig4}
\end{figure}

\begin{figure}[htp]
    \includegraphics[width=0.5\columnwidth]{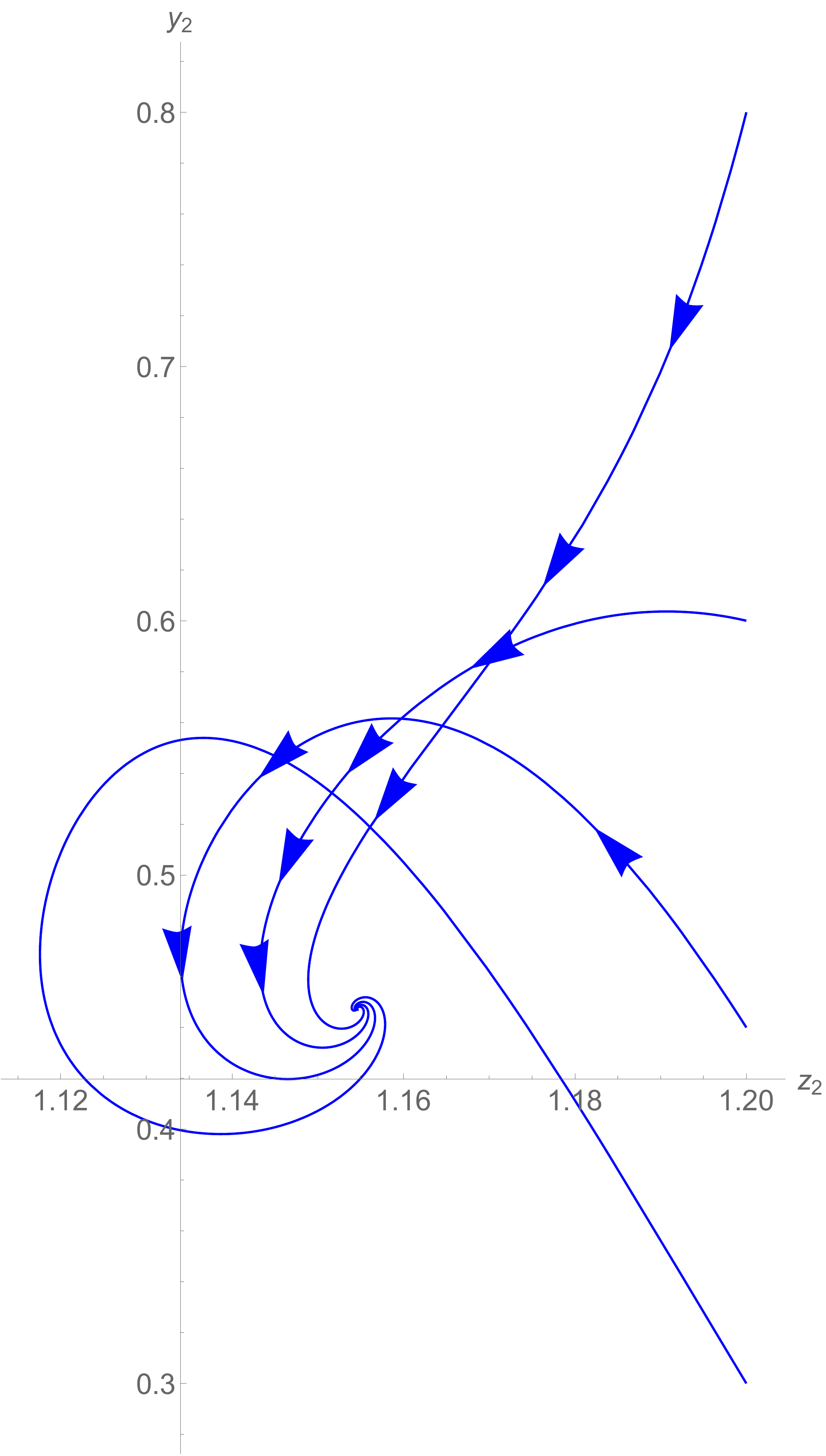}
    \caption{The evolution towards the critical point $P_{6+}$ ($\alpha_1= 10, \alpha_2= 12, \xi_1= 0.2, \xi_2 = - 0.1, w_m=0$).}
    \label{fig6bbb}
\end{figure}

\begin{figure}[htp]
    \includegraphics[width=0.7\columnwidth]{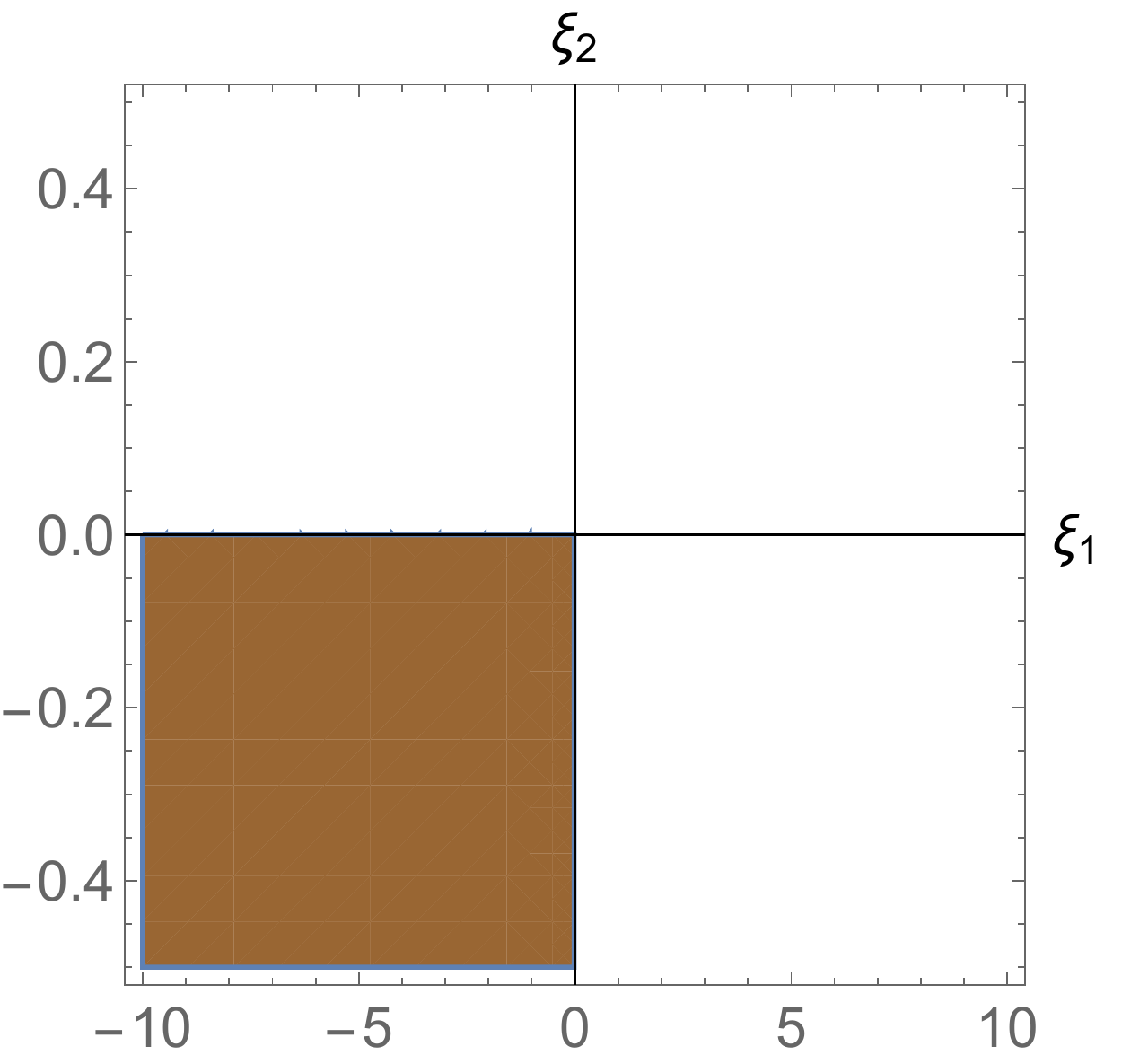}
    \caption{The non-exclusive saddle regions where the $P_{7+}$ critical point represents a saddle cosmological epoch ($\alpha_1=-0.5, \alpha_2=3, w_m=0, \xi_1 \in [-10,+10]$).}
    \label{fig5}
\end{figure} 

\begin{figure}[htp]
    \includegraphics[width=0.7\columnwidth]{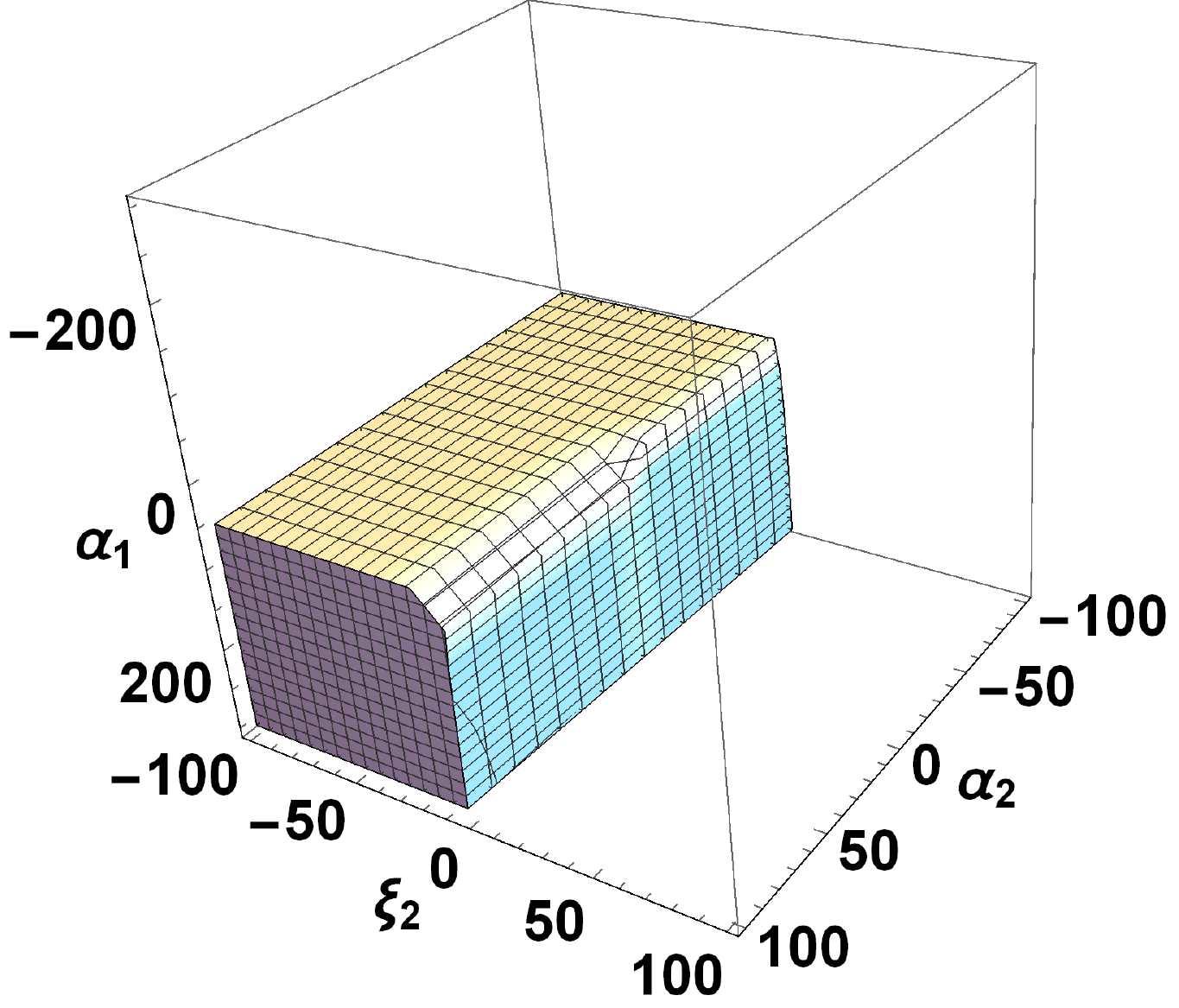}
    \caption{The non-exclusive saddle regions where the $P_{8}$ critical point represents a saddle cosmological epoch ($w_m=0, z_1=0, \xi_1=2, E_{P_{8}}[6]>0, E_{P_{8}}[2]<0$).}
    \label{fig6}
\end{figure}

\begin{figure}[htp]
    \includegraphics[width=0.6\columnwidth]{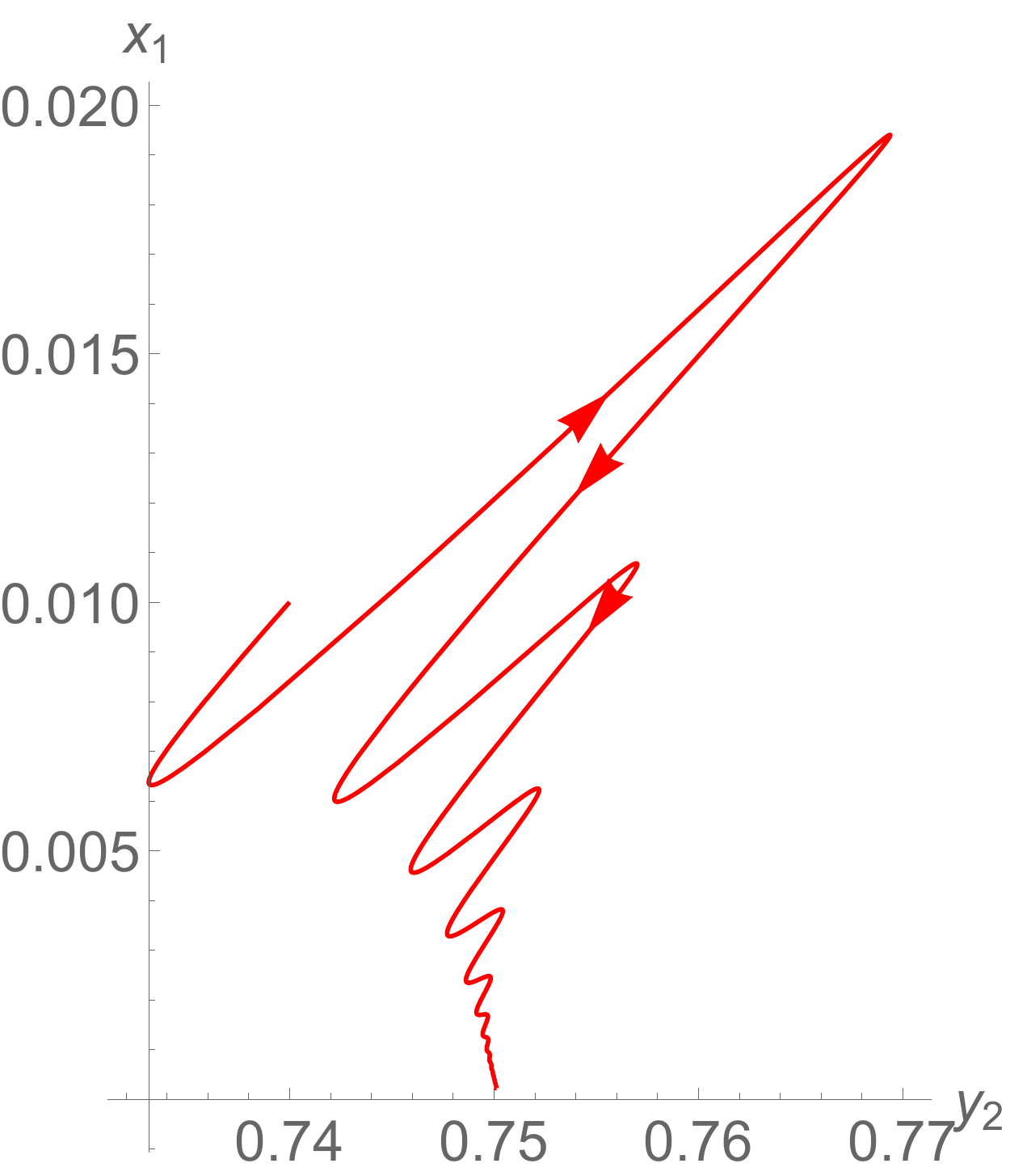}
    \caption{The evolution towards $P_8$ critical point ($\xi_1= 0.5,\xi_2= 10,\alpha_1= 30,\alpha_2= 3,w_m= 0$).}
    \label{fig8ev}
\end{figure}

\begin{figure}[htp]
    \includegraphics[width=0.6\columnwidth]{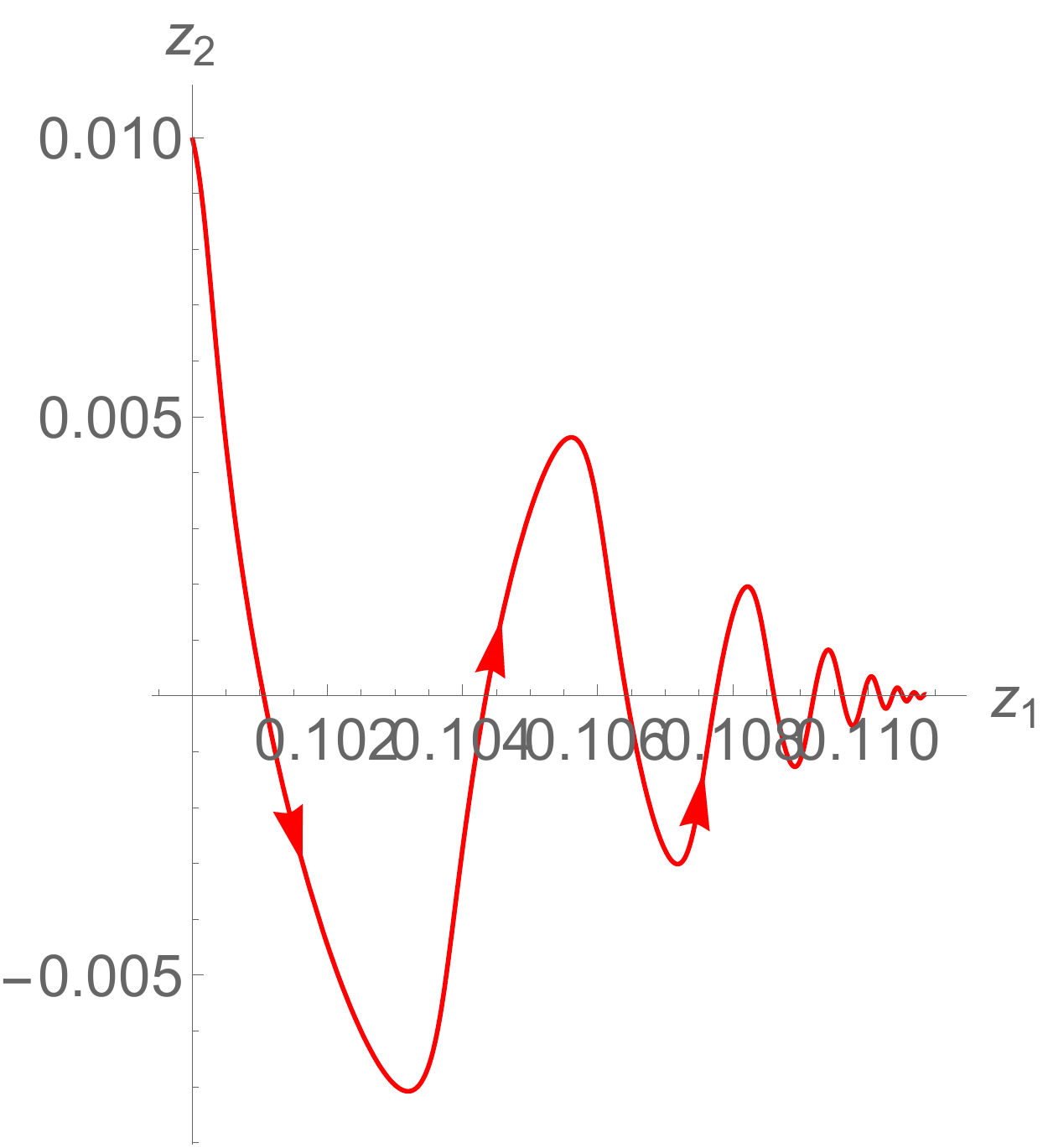}
    \caption{The evolution towards $P_8$ critical point in the $\{z_1,z_2\}$ variables for the same values of the parameters as in Fig.~\ref{fig8ev}.}
    \label{fig8ev2}
\end{figure}

\begin{figure}[htp]
    \includegraphics[width=0.7\columnwidth]{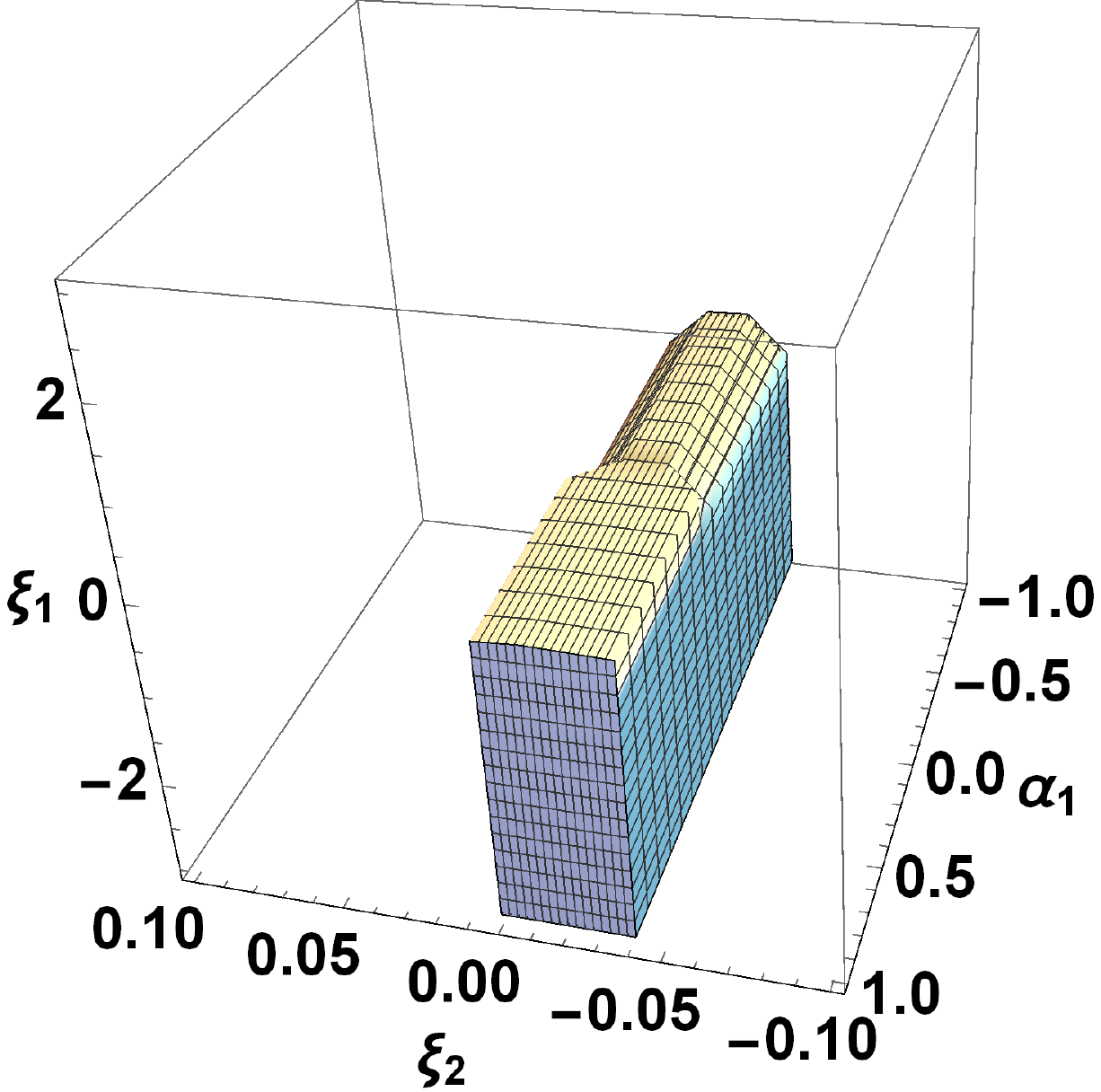}
    \caption{The non-exclusive saddle regions where the $P_{9}$ critical point represents a saddle cosmological epoch ($z_2=0, \alpha_2=1, E_{P_{9}}[5]>0, E_{P_{9}}[2]<0,  w_m=0$).}
    \label{fig7}
\end{figure}

\begin{figure}[htp]
    \includegraphics[width=0.7\columnwidth]{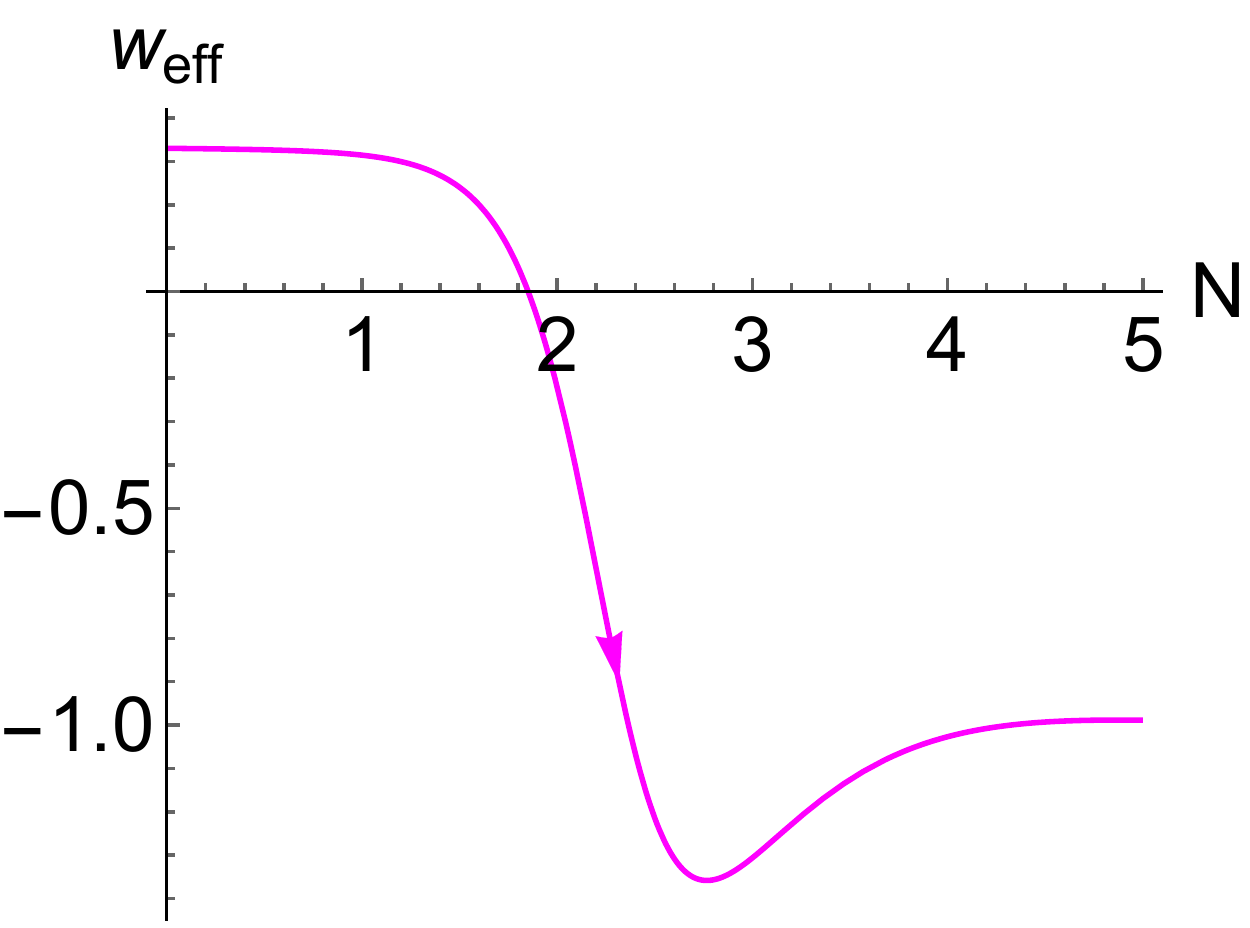}
    \caption{The variation of the effective equation of state towards a de--Sitter cosmological epoch in the case where $\xi_1= 0.5,\xi_2= 10,\alpha_1= 30,\alpha_2= 3,w_m= 0$.}
    \label{fig8}
\end{figure}

\par 
\section{Dynamical description of the model}
\label{sec:trei}
\par 
After having written the basic equations that describe the corresponding dark energy model, we shall try to investigate the dynamical properties of the cosmological scenario by making use of the linear stability theory. The dynamical analysis based on the linear stability theory represents an important tool used for investigating the physical characteristics of various scalar tensor theories of gravitation \cite{Bahamonde:2017ize}. For the specific cosmological scenario, we choose the following auxiliary variables \cite{HRYCYNA2010191}:
\begin{equation}
x_1=\frac{\dot{\phi}}{\sqrt{6}H},
\end{equation}
\begin{equation}
y_1=\frac{\sqrt{V_1(\phi)}}{\sqrt{3}H},
\end{equation}
\begin{equation}
z_1=\frac{\phi}{\sqrt{6}},
\end{equation}
\begin{equation}
\lambda_1=-\sqrt{6}\frac{1}{V_1(\phi)}\frac{dV_1(\phi)}{d \phi}
\end{equation}

\begin{equation}
x_2=\frac{\dot{\sigma}}{\sqrt{6}H},
\end{equation}
\begin{equation}
y_2=\frac{\sqrt{V_2(\sigma)}}{\sqrt{3}H},
\end{equation}
\begin{equation}
z_2=\frac{\sigma}{\sqrt{6}},
\end{equation}
\begin{equation}
\lambda_2=-\sqrt{6}\frac{1}{V_2(\sigma)}\frac{dV_2(\sigma)}{d \sigma}.
\end{equation}
\par
By introducing the specific variable $N=log(a)$ we can write the dynamics of the present cosmological model as an autonomous system of differential equations:
\begin{equation}
\label{afirst}
\frac{d x_1}{dN}=\frac{1}{\sqrt{6}} \frac{\ddot{\phi}}{H^2}-x_1 \frac{\dot{H}}{H^2},
\end{equation}
\begin{equation}
\frac{d y_1}{dN}=-\frac{\lambda_1}{2}x_1 y_1-y_1 \frac{\dot{H}}{H^2},
\end{equation}
\begin{equation}
\frac{d z_1}{dN}=x_1,
\end{equation}
\begin{equation}
\frac{d\lambda_1}{dN}=-\lambda_1^2 x_1 (\Gamma_1-1),
\end{equation}
\begin{equation}
\frac{d x_2}{dN}=\frac{1}{\sqrt{6}} \frac{\ddot{\sigma}}{H^2}-x_2 \frac{\dot{H}}{H^2},
\end{equation}
\begin{equation}
\frac{d y_2}{dN}=-\frac{\lambda_2}{2}x_2 y_2-y_2 \frac{\dot{H}}{H^2},
\end{equation}
\begin{equation}
\frac{d z_2}{dN}=x_2,
\end{equation}
\begin{equation}
\label{alast}
\frac{d\lambda_2}{dN}=-\lambda_2^2 x_2 (\Gamma_2-1),
\end{equation}
where $\Gamma_{i}$ ($i=1,2$) are defined as:
\begin{equation}
\Gamma_1=\frac{1}{\Big(\frac{dV_1(\phi)}{d\phi}\Big)^2}V_1(\phi)\frac{d^2 V_1(\phi)}{d \phi^2},
\end{equation}
\begin{equation}
\Gamma_2=\frac{1}{\Big(\frac{dV_2(\sigma)}{d\sigma}\Big)^2}V_2(\sigma)\frac{d^2 V_2(\sigma)}{d \sigma^2}.
\end{equation}
\par 
In what follows we shall assume a specific potential energy type where $V_i$ ($i=1,2$) have the form \cite{HRYCYNA2010191}:
\begin{equation}
V_1(\phi)=V_{10} exp\Big[-\frac{1}{6}\Big(\frac{\alpha_1}{2}\phi^2+\beta_1 \phi\Big)\Big],
\end{equation}
\begin{equation}
V_2(\sigma)=V_{20} exp\Big[-\frac{1}{6}\Big(\frac{\alpha_2}{2}\sigma^2+\beta_2 \sigma\Big)\Big].
\end{equation}
\par 
In the case where $\beta_i$ ($i=1,2$) are equal to zero then we can obtain different inter--relations between $z_i$ and $\lambda_i$ variables,
\begin{equation}
z_{i}(\lambda_{i})=\frac{\lambda_{i}}{\alpha_{i}},
\end{equation}
\begin{equation}
\Gamma_{i}=1-\frac{\alpha_{i}}{\lambda_{i}^2},
\end{equation}
reducing the dimension of the corresponding phase space from eight to six independent variables ($x_1,y_1,z_1,x_2,y_2,z_2$).
\par 
In the previous calculations, the ordinary differential system of equations is complete if we add the following identities which are deduced from the modified Friedmann eqs. and the Klein--Gordon relations:  
\onecolumngrid
\begin{multline}
\dot{H}=\frac{3 H^2}{2 \left(36 \xi _1^2 z_1^2-6 \xi _1 z_1^2-36 \xi _2^2 z_2^2+6 \xi _2 z_2^2+1\right)}(12 \xi _1 x_1 z_1 w_m-12 \xi _2 x_2 z_2 w_m+x_1^2 w_m-x_2^2 w_m+y_1^2 w_m+y_2^2 w_m+6 \xi _1 z_1^2 w_m
\\-6 \xi _2 z_2^2 w_m-w_m+4 \xi _1 x_1^2-4 \xi _2 x_2^2-4 \xi _1 x_1 z_1+4 \xi _2 x_2 z_2-x_1^2+x_2^2+2 \alpha _1 \xi _1 y_1^2 z_1^2+2 \alpha _2 \xi _2 y_2^2 z_2^2+y_1^2+y_2^2-48 \xi _1^2 z_1^2+48 \xi _2^2 z_2^2
\\+6 \xi _1 z_1^2-6 \xi _2 z_2^2-1),
\end{multline}

\begin{equation}
\ddot{\phi}=\sqrt{\frac{3}{2}} H^2 y_1^2 \lambda_1-3 \sqrt{6} H^2 x_1-12 \sqrt{6} H^2 \xi _1 z_1-6 \sqrt{6} \xi _1 z_1 \dot{H},
\end{equation}
\begin{equation}
\ddot{\sigma}=-\sqrt{\frac{3}{2}} H^2 y_2^2 \lambda_2-3 \sqrt{6} H^2 x_2-12 \sqrt{6} H^2 \xi _2 z_2-6 \sqrt{6} \xi _2 z_2 \dot{H},
\end{equation}
\begin{equation}
\Omega_m=1-\Omega_{\phi\sigma}=-12 \xi _1 x_1 z_1+12 \xi _2 x_2 z_2-x_1^2+x_2^2-y_1^2-y_2^2-6 \xi _1 z_1^2+6 \xi _2  z_2^2+1,
\end{equation}
\begin{multline}
w_{\bf{eff}}=-1-\frac{1}{36 \xi _1^2 z_1^2-6 \xi _1 z_1^2-36 \xi _2^2 z_2^2+6 \xi _2 z_2^2+1} \cdot \Big(12 \xi _1 x_1 z_1 w_m-12 \xi _2 x_2 z_2 w_m+x_1^2 w_m-x_2^2 w_m+y_1^2 w_m+y_2^2 w_m
\\+6 \xi _1 z_1^2 w_m-6 \xi _2 z_2^2 w_m-w_m+4 \xi _1 x_1^2-4 \xi _2 x_2^2-4 \xi _1 x_1 z_1+4 \xi _2 x_2 z_2-x_1^2+x_2^2+2 \text{$\alpha $1} \xi _1 y_1^2 z_1^2+2 \text{$\alpha $2} \xi _2 y_2^2 z_2^2+y_1^2+y_2^2-48 \xi _1^2 z_1^2
\\+48 \xi _2^2 z_2^2+6 \xi _1 z_1^2-6 \xi _2 z_2^2-1\Big).
\end{multline}
\twocolumngrid
\par
The critical points for the specific quintom scenario described by the action \eqref{actiune} are determined by setting the right hand sides of the equations \eqref{afirst}--\eqref{alast} to zero, considering only the relations for the ($x_1,y_1,z_1,x_2,y_2,z_2$) independent auxiliary variables, displayed in Table \ref{table:1}. In the following we shall analyze each critical point in detail, studying the fundamental properties from a physical and a dynamical point of view. For each critical point we have to take into consideration the acceptable physical existence conditions which require that the solutions are in the real phase space with non--zero denominators, and $y_{1,2}$ have positive real values, taking into account that for the location in the phase space structure all the expressions inside the square roots have to be positive. Due to the complexity of the phase space structure, we shall omit the presentation of the existence conditions for the critical points in our analysis.
\par 
In the dynamical system analysis presented in the present paper we have discussed the following types of critical points: stable, unstable, and saddle dynamical solutions. For the stable solutions any trajectory starting in a vicinity of the corresponding critical point and located in the attractor basin will lead to attaining the location of the dynamical solution in a given time. This type of solutions is characterized by the negativity of the real part of all the eigenvalues for the corresponding Jacobian evaluated in the specific solution. In a similar way, the unstable solutions are defined by the positivity of all the real parts of the resulting eigenvalues, characterized by the repelling of the trajectories in the phase space structure. The remaining class of critical points represents an intermediate type between the stable and unstable solutions, characterized by the positivity of some of the real parts of the eigenvalues, and the negativity of the real parts of different resulting eigenvalues. For further details related to the dynamical analysis the interested reader might consult Ref.~\cite{Bahamonde:2017ize}.
\par 
The first class of solutions $P_{1\pm}$ represents a critical line associated to a cosmological saddle scenario characterized by the domination of the dark energy component over the matter sector, where the dark energy mimics a radiation era. The eigenvalues for this critical line are the following:
\begin{equation}
E_{P_{1\pm}}=[0,2,2,-1,-1,1-3 w_m].
\end{equation}
In this case the kinetic and the potential energy terms of the quintom fields do not affect the location in the phase space structure and the dynamical features of the cosmological solutions, the critical line represent a saddle behavior independently to the values of various coupling parameters and constants. For these cosmological solutions we note an inter--relation between the value of the quintessence field $\phi$ embedded into the dynamical variable $z_1$, the value of the phantom field $\sigma$ represented by the $z_2$, and the two coupling coefficients $\xi_1$ and $\xi_2$.
\par 
The next class of dynamical solutions $P_{2\pm}$ represent a critical line where the auxiliary variable $z_2$ related to the value of the phantom field $\sigma$ has a real independent value. In this case the quintom fields are frozen, without any kinetic energy, while the potential energy is affected by the proportion between the curvature coupling coefficients $\xi_{1,2}$ and the corresponding potential energy parameters $\alpha_{1,2}$. For the location in the phase space structure the value of the quintessence field $\phi$ embedded into the $z_1$ variable is affected by the value of the phantom field $\sigma$ and all the remaining parameters for the present model, $\xi_{1,2}, \alpha_{1,2}$ which describe the curvature couplings and the potential energy strengths. At this critical line we observe the full domination of the quintom dark energy over the matter sector, the cosmological solution corresponds to a de--Sitter era where the quintom dark energy model behaves approximately as a cosmological constant. At this critical line we have obtained the following eigenvalues:
\begin{equation}
E_{P_{2\pm}}=[0,-3 \left(w_m+1\right), Q_3, Q_4, Q_5, Q_6].
\end{equation}
We note that the expressions for the eigenvalues $Q_3, Q_4, Q_5, Q_6$ are too complex to be written in the manuscript. Hence in what follows we shall rely only on numerical evaluations in order to explain properly the basic dynamical features at the corresponding cosmological solutions. For the $P_{2\pm}$ critical lines we note the existence of one zero eigenvalue which appears in any case, signalizing the limitation of the linear stability theory. Due to this, for these particular solutions we can study only the saddle dynamical behavior, while for a complete analysis a different approach should be considered, like the center manifold/Lyapunov method, or numerical evaluations, in the case of a more complex space of parameters and constants. Because of the high complexity of the phase space structure and eigenvalues the analysis is performed considering only the linear stability method which shall analyze only saddle dynamical behaviors for the $P_{2\pm}$ critical line. In the case of $P_{2+}$ critical line we have displayed in Fig.~\ref{fig1} some of the non--exclusive regions for the $\alpha_{1,2}$ parameters due to the existence conditions, while in Fig.~\ref{fig2} we have plotted the corresponding regions where the critical line $P_{2+}$ have a saddle dynamical behavior, assuming that some of the model's parameters are set ($\xi_1=-4, \xi_2=-1, z_2=10, w_m=0$). 
\par 
The next critical line denoted as $P_3$ represents a de-Sitter era with the domination of the dark energy component over the matter sector in terms of density parameters, characterized by a specific inter--relation between the potential energy terms. For this specific solution, we have obtained the following eigenvalues:
\begin{multline}
E_{P_3}=\Bigg[0,-3(w_m+1),\frac{1}{2}(-3 \pm \sqrt{9-48 \xi_1+2 \alpha_1 y_1^2}),
\\\frac{1}{2} \left(\pm\sqrt{-2 \alpha _2-48 \xi _2+2 \alpha _2 y_1^2+9}-3\right)\Bigg].
\end{multline}
We have displayed in Fig.~\ref{fig3} some of the non--exclusive regions where the critical line  $P_3$ represents a saddle cosmological epoch, by taking into account the existence conditions and the signs of the third and fourth eigenvalues. As in the previous case, due to the existence of one zero eigenvalue, we rely our analysis only  on linear stability theory, taking into account possible saddle regions.
\par 
The critical point $P_4$ represents the origin of the phase space, a cosmological solution characterized by the matter domination in terms of density parameters, while the effective equation of state is equal to the barotropic parameter $w_m$. At this point we have obtained the following eigenvalues:
\onecolumngrid
\begin{multline}
E_{P_{4}}=\Big[ \frac{3}{2} \left(w_m+1\right),\frac{3}{2} \left(w_m+1\right),
\frac{1}{4} \left(\pm\sqrt{18 \left(8 \xi _1-1\right) w_m+9 w_m^2-48 \xi _1+9}+3 w_m-3\right),
\\\frac{1}{4} \left(\pm\sqrt{18 \left(8 \xi _2-1\right) w_m+9 w_m^2-48 \xi _2+9}+3 w_m-3\right)\Big], 
\end{multline}
\twocolumngrid
showing that in the dust case $(w_m=0)$ the $P_4$ solution has a saddle dynamical behavior.
\par 
The next class of solutions, $P_{5\pm}$ represent particular cases of the $P_{1\pm}$ critical lines, a radiation dominated cosmological epoch which reduces to $P_{1\pm}$ if we set the auxiliary variable $z_2$ associated to the value of the phantom field $\sigma$ to zero. This type of solution can be further neglected in the analysis.
\par 
The $P_{6\pm}$ critical points describe a de-Sitter epoch characterized by the influence of the potential term for the phantom field, together with the corresponding value for $\sigma$. The location of the critical point in the phase space is affected mainly by the coupling constant of the phantom field $\xi_2$, and the $\alpha_2$ parameter which encodes the strength for the potential energy term. In our analysis we have obtained the following eigenvalues for the $P_{6+}$ solution:
\onecolumngrid
\begin{multline}
E_{P_{6+}}=\Big[0,-\frac{\sqrt{-8 \alpha _2 \left(42 \xi _2+1\right)-7 \alpha _2^2-48 \left(84 \xi _2^2+4 \xi _2-3\right)}+3 \alpha _2+72 \xi _2-12}{2 \left(\alpha _2+24 \xi _2-4\right)},
\\\frac{\sqrt{-8 \alpha _2 \left(42 \xi _2+1\right)-7 \alpha _2^2-48 \left(84 \xi _2^2+4 \xi _2-3\right)}-3 \alpha _2-72 \xi _2+12}{2 \left(\alpha _2+24 \xi _2-4\right)},
\\-\frac{\sqrt{3} \sqrt{-\left(16 \xi _1-3\right) \left(\alpha _2+24 \xi _2-4\right){}^2}+3 \alpha _2+72 \xi _2-12}{2 \left(\alpha _2+24 \xi _2-4\right)},
\\\frac{\sqrt{3} \sqrt{-\left(16 \xi _1-3\right) \left(\alpha _2+24 \xi _2-4\right){}^2}-3 \alpha _2-72 \xi _2+12}{2 \left(\alpha _2+24 \xi _2-4\right)},-3 \left(w_m+1\right)\Big].
\end{multline}
\twocolumngrid
In Fig.~\ref{fig4} we have displayed possible regions where the $P_{6+}$ critical point have a saddle dynamical behavior, in the dust case where $w_m=0$. The evolution towards $P_{6+}$ critical point is represented in Fig.~\ref{fig6bbb}, considering specific values of the parameters and different initial conditions.
\par 
For the $P_{7\pm}$ class of solutions we also have a domination of the dark energy field in terms of density parameters, a de--Sitter cosmological epoch influenced by the potential part of the quintessence field, together with its corresponding value. The location of the critical point in the phase space structure depends on the values of the $\xi_1$ and $\alpha_1$ parameters, which encodes the value of the non--minimal curvature coupling and the strength of the potential energy for the canonical field $\phi$. In this case we have obtained the following eigenvalues:
\onecolumngrid
\begin{multline}
E_{P_{7\pm}}=\Big[0,-\frac{\sqrt{8 \alpha _1 \left(42 \xi _1+1\right)-7 \alpha _1^2-48 \left(84 \xi _1^2+4 \xi _1-3\right)}+3 \alpha _1-72 \xi _1+12}{2 \left(\alpha _1-24 \xi _1+4\right)},
\\\frac{\sqrt{8 \alpha _1 \left(42 \xi _1+1\right)-7 \alpha _1^2-48 \left(84 \xi _1^2+4 \xi _1-3\right)}-3 \alpha _1+72 \xi _1-12}{2 \left(\alpha _1-24 \xi _1+4\right)},
\\-\frac{\sqrt{3} \sqrt{\left(16 \xi _2-3\right) \left(-\left(\alpha _1-24 \xi _1+4\right){}^2\right)}+3 \alpha _1-72 \xi _1+12}{2 \alpha _1-48 \xi _1+8},
\\\frac{\sqrt{3} \sqrt{\left(16 \xi _2-3\right) \left(-\left(\alpha _1-24 \xi _1+4\right){}^2\right)}-3 \alpha _1+72 \xi _1-12}{2 \left(\alpha _1-24 \xi _1+4\right)},-3 \left(w_m+1\right)
\Big].
\end{multline}
\twocolumngrid
Considering the dust case we have displayed in Fig.~\ref{fig5} a possible non--exclusive saddle region for the $P_{7+}$ critical point, showing the variation for the corresponding parameters.
\par 
The $P_{8}$ solution represent also a de--Sitter era where the two quintom fields are frozen, without any kinetic energy, only with non--negligible potential energy terms. For this particular solution we have a critical line where the auxiliary variable related to the value of the quintessence field $\phi$ encoded into $z_1$ is a real free parameter, affecting the potential energy terms, together with the values of the $\xi_1$ and $\alpha_1$ parameters. For this critical line we have obtained the following expression of the eigenvalues:
\onecolumngrid
\begin{multline}
E_{P_{8}}=\Big[
0,-3 \left(w_m+1\right),
\\-\frac{3 \alpha _1+18 \alpha _1 \xi _1 \left(6 \xi _1-1\right) z_1^2+\sqrt{3} \sqrt{\alpha _1^2 \left(6 \xi _1 \left(6 \xi _1-1\right) z_1^2+1\right) \left(96 \alpha _1 \xi _1^2 z_1^4+2 \xi _1 z_1^2 \left(-8 \alpha _1+150 \xi _1-9\right)+3\right)}}{2 \left(\alpha _1+6 \alpha _1 \xi _1 \left(6 \xi _1-1\right) z_1^2\right)},
\\
\frac{-3 \alpha _1-18 \alpha _1 \xi _1 \left(6 \xi _1-1\right) z_1^2+\sqrt{3} \sqrt{\alpha _1^2 \left(6 \xi _1 \left(6 \xi _1-1\right) z_1^2+1\right) \left(96 \alpha _1 \xi _1^2 z_1^4+2 \xi _1 z_1^2 \left(-8 \alpha _1+150 \xi _1-9\right)+3\right)}}{2 \left(\alpha _1+6 \alpha _1 \xi _1 \left(6 \xi _1-1\right) z_1^2\right)},
\\-\frac{3 \alpha _1+18 \alpha _1 \xi _1 \left(6 \xi _1-1\right) z_1^2+\sqrt{\alpha _1 \left(6 \xi _1 \left(6 \xi _1-1\right) z_1^2+1\right){}^2 \left(48 \alpha _2 \xi _1+\alpha _1 \left(-2 \alpha _2-48 \xi _2+9\right)+12 \alpha _1 \alpha _2 \xi _1 z_1^2\right)}}{2 \left(\alpha _1+6 \alpha _1 \xi _1 \left(6 \xi _1-1\right) z_1^2\right)},
\\\frac{-3 \alpha _1-18 \alpha _1 \xi _1 \left(6 \xi _1-1\right) z_1^2+\sqrt{\alpha _1 \left(6 \xi _1 \left(6 \xi _1-1\right) z_1^2+1\right){}^2 \left(48 \alpha _2 \xi _1+\alpha _1 \left(-2 \alpha _2-48 \xi _2+9\right)+12 \alpha _1 \alpha _2 \xi _1 z_1^2\right)}}{2 \left(\alpha _1+6 \alpha _1 \xi _1 \left(6 \xi _1-1\right) z_1^2\right)}
\Big].
\end{multline}
\twocolumngrid
For this critical line, we have shown a possible region where the solution have a saddle dynamical behavior displayed in Fig.~\ref{fig6}, a non--exclusive interval by considering $w_m=0, z_1=0, \xi_1=2$. Furthermore, the evolution towards $P_8$ critical point have been displayed in Figs.~\ref{fig8ev}, \ref{fig8ev2} for some values of the parameters and specific initial conditions.
\par 
The last dynamical solution $P_9$ has a similar behavior, an inter--relation between the potential energies of the quintom fields and $\xi_2$, $\alpha_2$ parameters. For this solution the auxiliary variable which encodes the value of the phantom field $z_2$ is a free parameter, affecting the location in the phase space structure and the corresponding physical features. The eigenvalues have the following expressions:
\onecolumngrid
\begin{multline}
E_{P_{9}}=\Big[0,-3 \left(w_m+1\right),
\\-\frac{-3 \alpha _2+18 \alpha _2 \xi _2 \left(6 \xi _2-1\right) z_2^2+\sqrt{3} \sqrt{\alpha _2^2 \left(6 \xi _2 \left(6 \xi _2-1\right) z_2^2-1\right) \left(96 \alpha _2 \xi _2^2 z_2^4+2 \xi _2 z_2^2 \left(8 \alpha _2+150 \xi _2-9\right)-3\right)}}{2 \alpha _2 \left(6 \xi _2 \left(6 \xi _2-1\right) z_2^2-1\right)},
\\\frac{3 \alpha _2-18 \alpha _2 \xi _2 \left(6 \xi _2-1\right) z_2^2+\sqrt{3} \sqrt{\alpha _2^2 \left(6 \xi _2 \left(6 \xi _2-1\right) z_2^2-1\right) \left(96 \alpha _2 \xi _2^2 z_2^4+2 \xi _2 z_2^2 \left(8 \alpha _2+150 \xi _2-9\right)-3\right)}}{2 \alpha _2 \left(6 \xi _2 \left(6 \xi _2-1\right) z_2^2-1\right)},
\\-\frac{-3 \alpha _2+18 \alpha _2 \xi _2 \left(6 \xi _2-1\right) z_2^2+\sqrt{\alpha _2 \left(6 \xi _2 \left(1-6 \xi _2\right) z_2^2+1\right){}^2 \left(\alpha _2 \left(2 \alpha _1-48 \xi _1+9\right)+48 \alpha _1 \xi _2+12 \alpha _1 \alpha _2 \xi _2 z_2^2\right)}}{2 \alpha _2 \left(6 \xi _2 \left(6 \xi _2-1\right) z_2^2-1\right)},
\\\frac{3 \alpha _2-18 \alpha _2 \xi _2 \left(6 \xi _2-1\right) z_2^2+\sqrt{\alpha _2 \left(6 \xi _2 \left(1-6 \xi _2\right) z_2^2+1\right){}^2 \left(\alpha _2 \left(2 \alpha _1-48 \xi _1+9\right)+48 \alpha _1 \xi _2+12 \alpha _1 \alpha _2 \xi _2 z_2^2\right)}}{2 \alpha _2 \left(6 \xi _2 \left(6 \xi _2-1\right) z_2^2-1\right)}\Big].
\end{multline}
\twocolumngrid
For the last solution we have displayed in Fig.~\ref{fig7} a non--exclusive three dimensional region where the dynamical features corresponds to a saddle behavior, by considering $z_2=0, \alpha_2=1$.

\section{Conclusions}
\label{sec:patru}
\par  

In this work we have studied a quintom cosmological model having a non-negligible non--minimal coupling with gravity through the scalar curvature. After presenting the modified Friedmann relations and the Klein--Gordon equations which describe the fundamental evolutionary aspects for the present cosmological scenario, we have analyzed the dynamical properties of the model by assuming the linear stability theory for a specific potential energy type. In this case we have assumed that the potential energy part in the corresponding action is represented by an exponential type squared, reducing the dimension of the phase space structure to six independent auxiliary variables due to specific inter--relations. By adopting the linear stability theory we have investigated the fundamental properties of the phase space structure, constraining from a dynamical point of view the corresponding parameters associated to the cosmological model. The investigation showed that the dynamical solutions corresponding to the critical points can explain various fundamental epochs in the current evolution of the Universe, including the radiation or matter dominated stages and the de--Sitter era, where the quintom model behaves as a cosmological constant. In this case each dynamical solution is investigated in detail, obtaining possible constraints for the model's parameters from a dynamical perspective. Note that in our analysis we have omitted the presentation of the existence conditions due to the high complexity of the logical expressions involved.
\par 
As can be noted from the analysis, the non--minimal curvature couplings $\xi_{1,2}$ affect the structure of the phase space and the dynamical features of the critical points, together with the values of the $\alpha_{1,2}$ parameters which encodes the strength of the potential energy type. The potential energy of the quintom scenario is a specific exponential case which enables us to reduce the dimension of the resulting phase space with two degrees of freedom. Analyzing the structure of the phase space and the location of the associated critical points, we have noticed that in this case the non--minimal coupling coefficients $\xi_{1,2}$ and the values of the $\alpha_{1,2}$ parameters affects the physical features involved and the corresponding dynamical effects. In this case we have obtained possible constraints for the coupling parameters $\xi_{1,2}$ and potential energy constants $\alpha_{1,2}$ from a physical and a dynamical point of view, associated to different physical features of the phase space. We have observed that all the cosmological solutions have a zero kinetic energy and can be regarded as frozen in time. The cosmological epochs in the phase space structure are associated to different dynamical solutions which can explain some of the evolutionary aspects related to the history of our Universe. To summarize, in the structure of the phase space we have obtained the following dynamical eras: radiation (described by the $P_{1, 5}$ critical points), matter domination (the origin of phase space, the $P_4$ solution), and de--Sitter (the remaining $P_{2, 3, 6, 7, 8, 9}$ cosmological solutions). 
\par 
In this context we have showed in Fig.~\ref{fig8} the dynamics of the effective equation of state in this model from an epoch where the evolution mimics a radiation era, passing through a transient matter dominated transition, finalizing in an asymptotic manner as a de Sitter stage, where the dark energy fluid behaves closely to the cosmological constant. The evolutionary aspects showed that the effective equation of state can exhibit phantom divide line crossing as a specific phenomena associated in general to quintom scenarios, appending a viable physical feature to the scenario. Finally we can note that the present cosmological model represents a possible extension to general relativity which can explain the existence of radiation, matter dominated epochs, and the current evolution closely to the cosmological constant, a feasible scenario which deserves further astrophysical investigations. 
\par 
In principle, the analysis described in the present paper is limited to the usage of linear stability theory, an important analytical tool considered in various scalar tensor theories. However, a more complete understanding of the dynamical features 
implies the consideration of various observational signatures, adding viable constraints to the present proposal. Hence, the present model in scalar tensor theories can be further studied in various cosmological applications. It is expected that the non-minimal couplings \cite{Geng:2015nnb} affects the gravitational interaction on local scales and can be used as a probe to study different aspects of scalar tensor theories, by considering different solar system constraints. For example, since the non-minimal couplings affect the gravitational interaction on local scales one can consider a study which takes into account possible observational signatures, further analyzing the confidence intervals for various associated parameters by taking into account different observational constraints.

\section{Acknowledgements}
The author would like to thank C.M. for support and suggestions. For the development of this project various analyses have been performed in Wolfram Mathematica \cite{Mathematica}.

\bibliography{apstemplate}
\bibliographystyle{apsrev}

\end{document}